\documentclass[conference]{IEEEtran}
\IEEEoverridecommandlockouts
\usepackage{tikz}

\newcommand\copyrighttext{%
  \footnotesize \textcopyright 2026 IEEE. Personal use of this material is permitted. 
  Permission from IEEE must be obtained for all other uses, in any current or future 
  media, including reprinting/republishing this material for advertising or promotional 
  purposes, creating new collective works, for resale or redistribution to servers or 
  lists, or reuse of any copyrighted component of this work in other works.}

\newcommand\copyrightnotice{%
\begin{tikzpicture}[remember picture,overlay]
\node[anchor=south,yshift=20pt] at (current page.south) {\parbox{\textwidth}{ \copyrighttext}};
\end{tikzpicture}%
}

\usepackage{algorithm}
\usepackage{algpseudocode}
\usepackage{cite}
\usepackage{float}
\usepackage{amsmath,amssymb,amsfonts,amsthm}
\usepackage{subcaption}
\usepackage{graphicx}
\usepackage{textcomp}
\usepackage{xcolor}
\usepackage{hyperref}
\usepackage{multirow,array,multicol}
\def\BibTeX{{\rm B\kern-.05em{\sc i\kern-.025em b}\kern-.08em
    T\kern-.1667em\lower.7ex\hbox{E}\kern-.125emX}}

\usepackage{mathtools}
\mathtoolsset{showonlyrefs}
\usepackage{bm,bbm}
\usepackage{xcolor}
\newtheorem{theorem}{Theorem}

\newtheorem{definition}{Definition}
\allowdisplaybreaks

\newcommand{\alg}{{\textsc{alg}}}
\newcommand{\ross}{{\textsc{ross}}}

\newcommand{\adv}{{\textsc{adv}}}
\newcommand{\opt}{{\textsc{opt}}}

\newcommand{\E}{\mathbbm{E}}
\newcommand{\p}{\mathbbm{P}}

\newcommand\red[1]{\textcolor{black!10!red}{#1}}

\makeatletter
\newcommand{\linebreakand}{%
  \end{@IEEEauthorhalign}
  \hfill\mbox{}\par
  \mbox{}\hfill\begin{@IEEEauthorhalign}
}
\makeatother

\makeatletter
\newenvironment{subroutine}[1][htb]{%
    \renewcommand{\ALG@name}{Sub-Routine}% Update algorithm name
   \begin{algorithm}[#1]%
  }{\end{algorithm}}
\makeatother

\begin{document}
% \iffalse
% \title{Robust Job Scheduling on Cloud Spot Instances: Optimal Guarantees using Randomization
% \thanks{*while author was working at Nokia Bell Labs, Murray Hill, NJ, USA.}
% }
% \fi

\title{Exploiting Spot Instances for Time-Critical Cloud Workloads Using Optimal Randomized Strategies\thanks{*while author was working at Nokia Bell Labs, Murray Hill, NJ, USA.}}

\author{
\IEEEauthorblockN{Neelkamal Bhuyan*}
\IEEEauthorblockA{\textit{Georgia Institute of Technology} \\
Atlanta, GA, USA \\
nbhuyan3@gatech.edu}
\and
\IEEEauthorblockN{Randeep Bhatia}
\IEEEauthorblockA{\textit{Nokia Bell Labs}\\
Murray Hill, NJ, USA \\
randeep.bhatia@nokia-bell-labs.com}
\linebreakand
\IEEEauthorblockN{Murali Kodialam}
\IEEEauthorblockA{\textit{Nokia Bell Labs}\\
Murray Hill, NJ, USA \\
murali.kodialam@nokia-bell-labs.com}
\and
\IEEEauthorblockN{TV Lakshman}
\IEEEauthorblockA{\textit{Nokia Bell Labs}\\
Murray Hill, NJ, USA \\
tv.lakshman@nokia-bell-labs.com}
}

\maketitle
\copyrightnotice
\begin{abstract}
This paper addresses the challenge of deadline-aware online scheduling for jobs in hybrid cloud environments, where jobs may run on either cost-effective but unreliable spot instances or more expensive on-demand instances, under hard deadlines. We first establish a fundamental limit for existing (predominantly-) deterministic policies, proving a worst-case competitive ratio of $\Omega(K)$, where $K$ is the cost ratio between on-demand and spot instances. We then present a novel \textit{randomized} scheduling algorithm, \ross{}, that achieves a provably optimal competitive ratio of $\sqrt{K}$ under reasonable deadlines, significantly improving upon existing approaches. 
%We further complement our theoretical findings with a matching lower bound for randomized policies, thereby establishing the optimality of \ross{}. 
Extensive evaluations on real-world trace data from Azure and AWS demonstrate that \ross{} effectively balances cost optimization and deadline guarantees, consistently outperforming the state-of-the-art by up to $30\%$ in cost savings, across diverse spot market conditions.

\end{abstract}


\section{Introduction}

As cloud computing becomes the default deployment platform for modern workloads \cite{reaufsm2023adoption,ahmad2024cloud,kutzias2019behind,precedence2024cloudmarket}, controlling operational costs has emerged as a critical challenge for both enterprises and start-ups. One of the dominant contributors to cloud costs is compute infrastructure, typically provisioned through on-demand instances that offer guaranteed availability but at a high price. To improve resource utilization at the  and reduce operating costs, cloud providers like Amazon Web Services (AWS), Microsoft Azure, Google Cloud Platform and Oracle offer spot instances \cite{AWS_EC2_prices,azure_spots,GCP,oracle_spots}—unused capacity sold at steep discounts (often 3–10× cheaper) \cite{295489,javadi2011statistical,singh2015dynamic} —but with the caveat that they can be revoked at short notice. This cost-performance trade-off has made spot instances increasingly attractive for a range of compute-intensive workloads, including machine learning training, data analytics, and scientific computing. Many of these applications incorporate fault-tolerant mechanisms such as checkpointing or restart-on-failure to recover from preemptions, thus enabling them to leverage the cost savings of spot instances effectively.

However, not all applications can afford the latency and uncertainty introduced by spot instance revocations. In particular, delay-sensitive workloads involving latency-bound inference pipelines in recommendation systems\cite{cao2017deep,zhang2019deep,kang2020efficient}, IoT \cite{yu2020low,guo2023real}, edge computing \cite{11181549,bhuyan2022multi,bhuyan2022provable} and video computing \cite{zhang2018delay}—require strict adherence to deadlines and consistent computational performance. For these applications, relying solely on spot instances can result in missed deadlines, degraded service, or economic loss. As a result, practitioners often resort to expensive on-demand instances to ensure reliability, sacrificing cost efficiency for predictability. This creates a fundamental tension between minimizing cost and meeting delay guarantees. While heuristic policies such as "greedy" switching—waiting on spot until the remaining time equals job duration—offer simple solutions, they are often suboptimal. There remains a pressing need for principled strategies that navigate this trade-off effectively, enabling the use of low-cost spot resources without violating delay constraints.  

The starting point of our investigation is \cite{295489}, 
which addresses the challenge of minimizing cloud computing costs while meeting job deadlines by leveraging low-cost but unreliable spot instances. The authors analyze spot instance behavior—availability, pricing, and lifetimes—across cloud regions to inform the design of robust scheduling policies. 
% They develop a theoretical framework to evaluate the worst- and average-case performance of various policies, including a baseline greedy strategy. 
Their core contribution is Uniform Progress, a simple, deterministic, parameter-free scheduling policy that distributes job execution evenly over time, and performing well against the greedy baseline  on real-world spot instance traces. 
% Through simulations on real-world spot instance traces and evaluations on diverse workloads, the authors demonstrate that Uniform Progress achieves 27–84\% cost savings while meeting deadlines.
The worst-case performance of the Uniform Progress algorithm, however, is noted to scale linearly with the cost ratio $K$, defined as the ratio between the on-demand instance cost and the spot instance cost.

%\subsection{Using Randomization to Improve Performance}
It is well established that randomization significantly enhances performance in online optimization~\cite{Albers2003Online}. By incorporating randomized decision rules, algorithms can hedge against worst-case inputs and mitigate the limitations of deterministic approaches~\cite{bhuyan2026scale,7524551,pmlr-v235-bhuyan24a,bhuyan2025estimate}. In many classical settings, such as paging, scheduling, and resource allocation, randomized algorithms achieve provably better competitive ratios ~\cite{Blum1994UniformMTS,bar2002minimizing,khanafer2013constrained} than any deterministic algorithm, making them a powerful tool in designing robust online strategies.

%\subsection{Randomized Online Spot Scheduler (\ross{})}
Building on this principle, we propose \ross{} (Randomized Online Spot Scheduler), a novel online algorithm designed for scheduling deadline-constrained jobs on cloud infrastructure with both spot and on-demand instances. 
%\ross{} aims to meet strict job deadlines while minimizing overall cost. 
The algorithm operates in phases: it starts with a warm-up period—either Greedy or Uniform—to accumulate partial job progress, and then invokes a randomized scheduling mechanism that reserves an on-demand instance for an optimal interval length, with a fuzzy start point to average out unfavorable spot availability.
% This randomization enables \ross{} to intelligently exploit the cost advantage of spot instances without risking deadline violations.
By balancing exploration and exploitation, \ross{} guarantees deadline adherence while achieving substantial cost reductions compared to the state-of-the-art Uniform Progress \cite{295489}.

% \subsection{Our Contributions}
% We study the problem of online scheduling of delay-sensitive jobs over spot and on-demand cloud instances, focusing on minimizing cost under a hard deadline constraint. 
\noindent \textit{Main Contributions:} In this work, we establish fundamental results for cost-efficient execution of deadline-constrained jobs on cloud spot instances:
\begin{itemize}
    \item \textbf{Fundamental Limit for Deterministic Policies:} We prove a worst-case lower bound showing that \emph{any deterministic online scheduling policy} must incur a competitive ratio of at least $\Omega(K)$, where $K$ is the cost ratio between on-demand and spot instances (Theorem~1).

    \item \textbf{Provable Competitive Ratio of \ross{}:} We analyze \ross{} and show that it achieves a competitive ratio of
    \[
    \mathrm{CR}_{\ross{}} =
    \begin{cases}
    \sqrt{K} & \text{if } D \geq \frac{1 + 2\sqrt{K}}{1 + \sqrt{K}} L \\
    1 + (K - 1) \left( 2 - \frac{D}{L} \right) & \text{otherwise}
    \end{cases}
    \]
    where $L$ is the job length and $D$ the deadline (Theorem~2). The competitive ratio of ROSS is $\sqrt{K}$ as opposed to $K$ for uniform progress proposed in \cite{295489} when the deadlines are not very tight. 
    
    \item We also prove a matching lower bound for all randomized policies (Theorem~3) and thus {\bf \ross{} is optimal}.

    \item \textbf{Empirical Evaluation with Real Spot Traces:} We evaluate \ross{} on publicly available datasets: SpotLake Archive \cite{lee2022spotlake} and SkyPilot traces \cite{295489} from Azure and AWS, under varying deadline ratios ($L/D$) and cost ratios ($K$). Our results show:
    \begin{itemize}
     \item Under \textbf{loose deadlines}, \ross{} significantly outperforms both Greedy and (state-of-the-art) Uniform Progress (by 30\%) by effectively utilizing slack to lower cost, avoiding overly conservative behavior.
    \item For \textbf{tight deadlines}, \ross{} with Uniform Warm-up matches the reliability of Uniform Progress, while exploiting scenarios with favorable spot availability.
    \item Across diverse spot availability traces, \ross{} consistently reduces costs compared to relying solely on on-demand instances, and closely approaches the performance of the offline optimal.
\end{itemize}
\end{itemize}
\vspace{-5pt}
\section{Related Work}
Amazon Web Services (AWS) introduced spot instances in 2009, using a bidding system to allocate unused cloud capacity~\cite{AWS_EC2_prices}. Early research explored various pricing strategies and bidding approaches for these spot markets~\cite{agmon2013deconstructing, javadi2011statistical, wang2013present, song2012optimal, singh2015dynamic, tang2012towards}, alongside efforts in scheduling under price uncertainty~\cite{menache2014demand, varshney2018autobot, poola2014fault, zafer2012optimal, song2012optimal}. However, as major providers transitioned to more predictable models—such as AWS EC2's stabilized spot pricing~\cite{AWS_EC2_prices}, Google Cloud Platform's (GCP) fixed 30-day rates~\cite{GCP}, Oracle Cloud's flat 50\% discount for preemptible instances~\cite{oracle_spots}, and Azure's regionally stable pricing~\cite{azure_spots}—traditional bidding-based scheduling lost its relevance. This evolution shifted research emphasis toward the \emph{availability} and \emph{reliability} of spot instance offerings.

Recent works have concentrated on collecting and statistically analyzing availability data across providers~\cite{ kadupitige2020modeling, 10.1145/3589334.3645548}. Growing attention has also been given to scheduling algorithms that leverage machine and deep learning to predict future spot availability, particularly within hybrid spot/on-demand resource contexts~\cite{harlap2018tributary, yang2022spot, yang2023snape}, and to workflow scheduling over such instances~\cite{taghavi2023cost, 9893047}. Nevertheless, many studies still lack robust consideration of deadline constraints. Only very recent approaches (e.g.,~\cite{9640599,295489,11038902,bhuyan2026opportunisticschedulingoptimalspot}) are beginning to address deadline-aware, cost-efficient scheduling by combining predictive models with dynamic resource optimization. 
%As spot markets stabilize and interruptions become more predictable, the active frontier of research is now defined by (i) statistical characterization of interruptions and reliability, and (ii) the design of intelligent, deadline-constrained scheduling policies informed by machine learning to maximize both cost savings and practical usability.
\vspace{-5pt}
\section{System Model}
We model job execution as an online scheduling problem where a single computational job must be completed before a deadline using cloud resources. The scheduler (also referred to as the “player” or “user”) has access to two types of instances offered by the environment (or “adversary”):

\begin{itemize}
\item \textbf{Spot instances}, which cost $1$ per unit time but whose availability is controlled by the environment and is \emph{not known in advance} to the scheduler.
\item \textbf{On-demand instances}, which are always available but cost $K > 1$ per unit time.
\end{itemize}

Both instance types execute the job at the same rate. To complete the job, a total of $L$ units of compute time must be accumulated across rented instances, and this must occur within a fixed time budget of $D \geq L$ units. The scheduler’s challenge is to make irrevocable decisions without knowledge of future spot instance availability, while minimizing total cost and ensuring the job finishes before the deadline.
We define $\phi(t)$ as the cumulative amount of computation completed by time $t$. Initially, $\phi(0) = 0$, and upon job completion at some time $C \leq D$, we have $\phi(C) = L$. 

While our theoretical analysis follows prior work \cite{295489,11038902} in omitting the negligible delays that occur when jobs switch between idle state and on-demand/spot instances, we explicitly account for these delays in our experimental evaluation (Section \ref{sims}) to ensure accurate real-world results.

\subsection{Competitive Analysis}
To evaluate the performance of an online scheduler in the presence of uncertainty, we adopt the standard metric of \emph{competitive analysis}. This framework compares the cost incurred by an online algorithm—which must act without knowledge of future spot availability—with that of an ideal offline policy that has complete knowledge of the future. The following definition formalizes this comparison through the notion of competitive ratio.
\begin{definition}
    The \emph{competitive ratio} of an online algorithm \alg{} is defined as the worst-case ratio between its expected cost and the cost of the optimal offline policy \opt{}, which has full knowledge of future spot instance availability:
    \vspace{-3pt}
    \begin{align*}
        CR_{\alg}(K) = \frac{\mathbb{E}[\text{Cost}_{\alg}(K)]}{\text{Cost}_{\opt}(K)}
    \end{align*}
    \vspace{-15pt}
\end{definition}
Here, $\text{Cost}_{\alg}(K)$ is the total cost incurred by the online algorithm when the on-demand cost is $K$ per unit time, and $\text{Cost}_{\opt}(K)$ is the minimum cost by an offline policy with complete foresight. The expectation $\mathbb{E}[\cdot]$ is taken over the possible internal randomness of the algorithm.
%, as randomized strategies are essential for achieving robustness against worst-case instance availability patterns. 
\subsection{Impact of Deadlines on the Scheduling Policy}
A central concept in our analysis is the notion of \emph{slack}, which quantifies the scheduler's flexibility at any point in time. Formally, slack $s(t)$ is defined as the difference between the remaining time until the deadline and the amount of compute still required to finish the job, that is,\vspace{-2pt}
\begin{align}
    s_{\alg}(t) = (D-t)-(L-\phi_{\alg}(t)).
\end{align}
When this slack reaches zero, the system enters a critical phase of execution where any further idling risks violating the deadline constraint. We define the \emph{point of no return} for a scheduling algorithm $ALG$, denoted by $T^{\mathrm{NR}}_{\alg}$, as the first time this slack vanishes, that is,
\vspace{-2pt}
\begin{align}
    T^{NR}_{\alg} = \min\{t\in [0,D]: s_{\alg}(t) = 0\}
\end{align}
At this point, the scheduler must continuously allocate compute resources until the job is completed. In particular, if no spot instance is available at $T^{\mathrm{NR}}_{\alg}$, the scheduler must resort to an on-demand instance for the remainder of the job in order to meet the deadline. The hindsight optimal \opt{} knows the adversary beforehand. So, we will often be referring to both as the same entity.
\vspace{-5pt}
\subsection{Deterministic v/s Randomized Algorithms}
Previous work \cite{295489} proposes algorithms that are deterministic with guarantees scaling linearly with $K$ (Theorem 1 and 2 in \cite{295489}). We provably generalize this limitation to the entire class of deterministic policies:  
\begin{theorem}\label{thm:det_policies}
    Any online scheduling policy taking deterministic decisions (idle/on-demand rental/spot rental) has a worst-case competitive ratio of $\Omega(K)$.
\end{theorem}
\vspace{-4pt}
The proof of this negative result is discussed in Section \ref{sec:proof}. It hinges on the adversary's ability to tailor the spot instance availability to the scheduler’s deterministic decisions. 
% The fundamental limitation of deterministic decisions: without randomness, the adversary can always orchestrate worst-case scenarios.
Consequently, this motivates the need for randomized algorithms, which can hedge against such adversarial patterns and offer provably better guarantees in expectation.

% Do you want to change the title of the next section to something else? We already used it once in the intro
\section{Randomized Optimal Spot Processing}
\label{sec:ross}
\begin{figure}
    \centering
    \includegraphics[width=\columnwidth]{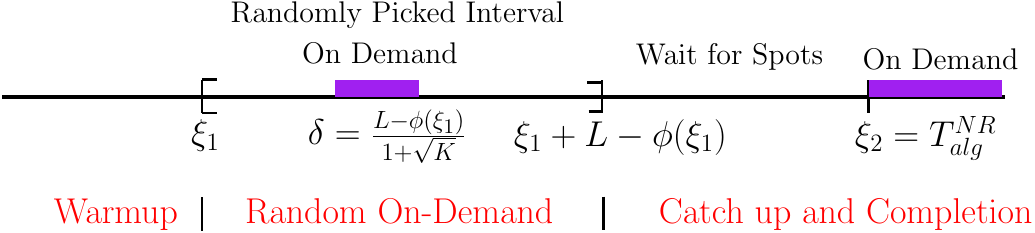}
    \caption{Schematic of \ross{}}
    \label{fig:ross_description}
    \vspace{-5pt}
\end{figure}

To overcome the inherent limitations of deterministic online scheduling, we introduce \ross{} (\textbf{R}andomized \textbf{O}nline \textbf{S}pot \textbf{S}cheduler), a randomized online algorithm designed to minimize job cost under deadline constraints. The key idea is to utilize early slack (when deadlines are loose) for making compute progress through randomized rental of on-demand instances.
Randomization allows \ross{} to hedge against worst-case spot instance placement along the horizon $[0,D]$ while provably maintaining deadline guarantees.
At a high level, \ross{} proceeds in three phases as shown in Figure \ref{fig:ross_description}:
\begin{enumerate}
    \item \textbf{Warm-up Phase:} As long as the running deadline-to-compute ratio $\left(\frac{D-t}{L - \phi(t)}\right)$ is \emph{above} the critical threshold $\frac{1 + 2\sqrt{K}}{1 + \sqrt{K}}$, the algorithm uses a \textit{warm-up} strategy to accumulate partial job progress. The warm-up policy can be either \textit{greedy} or \textit{uniform}, chosen via input parameter $w$. Greedy warm-up uses on-demand instances more aggressively, while uniform warm-up spreads work more evenly over time.
    \item \textbf{Randomized On-demand Injection:} Once the system reaches the critical ratio, the algorithm records the current time $\xi_1$, computes a randomized duration $\delta = \frac{L - \phi(\xi_1)}{1 + \sqrt{K}}$, and selects a random interval $I(\delta) \subseteq [\xi_1, \xi_1 + L-\phi(\xi_1)]$ of length $\delta$. During this interval, \ross{} rents on-demand instances to guarantee progress. Outside of this interval, it opportunistically uses spot instances (or idle if unavailable).
    \item \textbf{Catch-up and Completion:} After time $\xi_1 + L-\phi(\xi_1)$, if the job is still incomplete, the algorithm continues using spot instances whenever available, idling otherwise, until the point at which no slack remains. At this final point $\xi_2$, \ross{} uses on-demand instances to complete the remaining compute, ensuring deadline satisfaction.
\end{enumerate}
%\vspace{-50pt}
By strategically injecting a small randomized dose of on-demand usage during periods of slack, \ross{} balances deadline guarantees with cost savings, adapting to both tight and loose deadlines without requiring prior knowledge of future spot availability. A formal description of the algorithm is given in Algorithm~\ref{alg:ross_base}, with the warm-up phase in Sub-Routine~\ref{alg:ross_warm_up}.

We now turn to analyzing the performance of \ross{}. The following theorem characterizes its competitive ratio across all deadline regimes, showing that \ross{} achieves a provably sublinear competitive ratio of $\sqrt{K}$ when the deadline is not too tight compared a competitive ratio of $K$ for  the uniform progress algorithm in \cite{295489}. 

\begin{theorem}
    The competitive ratio of \ross{} for any compute and deadline $0\leq L \leq D$ is
    \begin{align*}
    CR_{\ross} =
    \begin{cases}
        \sqrt{K} & D \geq \frac{1 + 2\sqrt{K}}{1 + \sqrt{K}}L \\
        1 + (K - 1)\left(2 - \frac{D}{L}\right) & D \leq \frac{1 + 2\sqrt{K}}{1 + \sqrt{K}}L.
    \end{cases}
    \end{align*}
    %with Phase 1 of length $z^* = \min(L, D - L + \delta)$ and random on-demand rental of length $\delta^* = \frac{L}{\sqrt{K} + 1}$.
\end{theorem}

The proof of the above performance guarantee is discussed in detail in Section \ref{sec:proof}. The competitive ratio derived above not only demonstrates the efficiency of \ross{} across a wide range of deadline regimes, but it also turns out that this cannot be improved. In the next result, we establish a fundamental lower bound on the performance of any online scheduling algorithm, showing that no strategy—deterministic or randomized—can achieve a better competitive ratio than that attained by \ross{}. This establishes the optimality of \ross{} in terms of worst-case cost guarantees.

%\begin{minipage}{\columnwidth}
\begin{algorithm}[H]
\caption{\ross: \textbf{R}andomized \textbf{O}nline \textbf{S}pot \textbf{S}cheduler}
\label{alg:ross_base}
\begin{algorithmic}[1]
\State \textbf{Input:} Deadline $D$, Compute length $L$, Cost ratio $K$
\State \textbf{Initialize:} $t = 0$, $w \in \{\text{greedy}, \text{uniform}\}$, $\xi_1,\xi_2=0$, $\delta = 0$, $\phi(t)$ with $\phi(0) \gets 0$
\While{$\frac{D - t}{L - \phi(t)} \leq \frac{1 + 2\sqrt{K}}{1 + \sqrt{K}}$}
    \State $\phi(t) \gets$ \textsc{WarmUp}($w$)
\EndWhile
\State $\xi_1 \gets t$
\State $\delta \gets \frac{L - \phi(\xi_1)}{1 + \sqrt{K}}$
\State $I(\delta) \gets$ Random interval in $[\xi_1, \xi_1 + L-\phi(\xi_1)]$ of size $\delta$
\While{$t \leq \xi_1 + L-\phi(\xi_1)$}
    \If{$t \in I(\delta)$}
        \State $\phi(t) \gets$ Rent on-demand instance
    \Else
        \State $\phi(t) \gets$ Rent available spot instance, o/w idle
    \EndIf
\EndWhile
\While{$(D - t) - (L - \phi(t)) > 0$}
    \State $\phi(t) \gets$ Rent available spot instance, o/w idle
\EndWhile
\State $\xi_2 \gets t$
\State Rent on-demand for remaining compute $(L - \phi(\xi_2))$
\end{algorithmic}
\end{algorithm}
%\end{minipage}
\vspace{-15pt}
\setcounter{algorithm}{0}
%\begin{minipage}{\columnwidth}
\begin{subroutine}[H]
\caption{Warm-Up Policy}
\label{alg:ross_warm_up}
\begin{algorithmic}[1]
\State \textbf{Input:} $w \in \{\text{greedy}, \text{uniform}\}$
%\State \textbf{Initialize:} Expected Progress $ep(t)$ with $ep(0) \gets 0$
\If{$w = \text{greedy}$}
    \State $\phi(t) \gets$ Rent available spot instance, o/w on-demand
\Else
    \If{$\phi(t) < \frac{L}{D}t$}
        \State $\phi(t) \gets$ Rent available spot instance, o/w on-demand
    \EndIf
\EndIf
\end{algorithmic}
\end{subroutine}
%\end{minipage}

\subsection{Optimality of \ross{}}
\noindent In this section we present the lower bound for the expected competitive ratio of {\em any} online scheduling algorithm.
\begin{theorem}\label{thm: lower_bound}
    The competitive ratio of any online algorithm \alg{} is lower bounded as
    \begin{align*}
    CR_{\alg} \geq \begin{cases}
            \sqrt{K} & D\geq \frac{1+2\sqrt{K}}{1+\sqrt{K}}L\\
            1+(K-1)\left(2-\frac{D}{L}\right) & D\leq \frac{1+2\sqrt{K}}{1+\sqrt{K}}L
        \end{cases}
    \end{align*}
\end{theorem}
The proof of the lower bound is given in Section \ref{sec:proof}. Theorem~\ref{thm: lower_bound} confirms that the competitive ratio achieved by \ross{} is not only favorable but also optimal for online scheduling under deadline constraints. With the theoretical performance of \ross{} matching the fundamental lower bound, we now turn to empirical validation. In the next section, we evaluate the practical effectiveness of \ross{} by comparing it against existing policies across diverse real-world spot instance traces and cost-delay scenarios.

\section{Simulations}\label{sims}
To further highlight the improvements \ross{} brings over the state-of-the-art scheduling policies, we compare \ross{} against the Uniform Progress policy \cite{295489}, greedy scheduling and the optimal hindsight scheduling policy \opt{} across various spot instance availability traces and environment settings.

\subsection{Set-up}
For a certain availability trace, we aim to measure the performance of four policies: (i) \ross{} with Greedy Warm-up, (ii) \ross{} with Uniform Warm-up, (iii) Uniform Progress and, (iv) naive greedy scheduling. We compare these policies under various settings of job length $L$, deadline $D$ and cost ratio $K$. Specifically, we consider $L/D$ to vary between a high of $0.9$ (strictest deadline) and a low value equal to the fraction of spots available over the horizon (loosest deadline). We consider $K \in (1,10]$, in accordance to practically observed values \cite{295489,singh2015dynamic,javadi2011statistical}.

In practice, jobs face change-over delays when switching states (idle/spot/on-demand). To take these into account, we consider the same average change-over delay used in \cite{295489}, that is $\approx 1\%$ of the compute length $L.$ We compare the policies over two metrics:
\begin{enumerate}
    \item Cost Savings (\%): This is \% cost benefit any \alg{} has over running purely on-demand. The \textit{higher} this metric is, the better.
    \item Overhead to \opt{} (\%): This is the \% extra cost \alg{} incurs in comparison to using all available spot instances over the horizon. For any \alg{}, this metric scales with its competitive ratio $CR_{\alg}$, hence, \textit{lower} is  better.
\end{enumerate}

\subsection{Datasets}
Our evaluation leverages all publicly available spot instance availability datasets, spanning major cloud providers. Specifically, we use data from Microsoft Azure and AWS EC2 spot VMs. Google Cloud Platform (GCP) does not release instance availability traces, and is therefore excluded from our study~\cite{lee2022spotlake}.

\paragraph{ \bf Spotlake Archive} The Spotlake archive~\cite{lee2022spotlake} maintains a continuously updated repository of spot instance availability data for both AWS and Azure. Each record corresponds to a specific VM type, defined by its provider, geographic region, and availability zone. Metadata includes estimated cost ratios and qualitative availability labels (low, medium, high). For simulation purposes, we treat a spot instance as available only when its availability label is \textit{high}; otherwise, we consider it unavailable. We conduct experiments using both AWS and Azure availability traces from this archive.

\paragraph{ \bf Skypilot Traces} To enable direct comparison with prior work, we use the AWS EC2 spot instance traces from~\cite{295489}, which were also used to evaluate the Uniform Progress policy. These consist of two trace types: (i) \textit{availability} traces, in which VMs are pinged every 10 minutes but immediately shut down, providing long-term availability snapshots; and (ii) \textit{preemption} traces, in which spot VMs are run until forcibly preempted, offering high-resolution insight into actual lifetimes.

\subsection{Empirical Findings}
We evaluate \ross{} across four distinct types of spot instance traces: (i) Spotlake Azure traces, (ii) Spotlake AWS traces, (iii) Skypilot AWS availability traces, and (iv) Skypilot AWS preemption traces. In each experiment, we examine two key trends: (a) percentage cost savings as the deadline becomes tighter $(L/D) \uparrow$, and (b) percentage cost gap to the offline optimal (\opt{}) as the cost ratio $K$ increases. These trends provide insight into the real-world practicality of each policy and reflect the theoretical benchmarks established in Section~\ref{sec:ross}.

\begin{figure}[t]
\centering
    \begin{subfigure}[b]{\columnwidth}
        \centering
        \includegraphics[width=\textwidth]{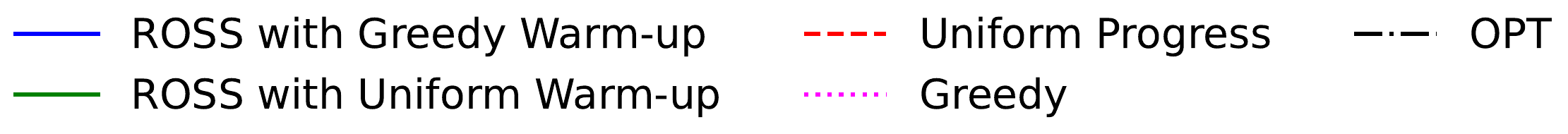}
    \end{subfigure}
    \begin{subfigure}[b]{\columnwidth}
        \centering
        \includegraphics[width=\textwidth]{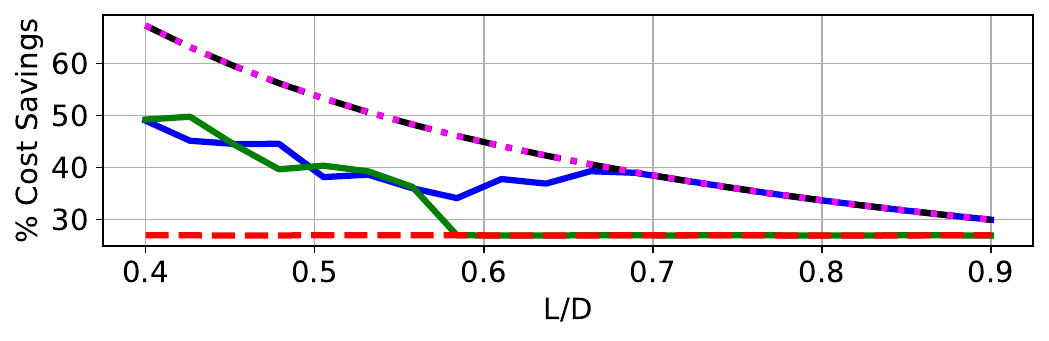}
        %\subcaption{Savings over on-demand}
        \label{fig:exp_light}
    \end{subfigure}
    \vspace{-25pt}
    % \begin{subfigure}[b]{0.5\columnwidth}
    %     \centering
    %     \includegraphics[width=\textwidth]{us-east-1f_v100_1/jf_0.5.pdf}
    %     \subcaption{Cost difference to Omniscient for $D=2\times L$}
    %     \label{fig:exp_overload}
    % \end{subfigure}%
    \\
    \begin{subfigure}[b]{0.5\columnwidth}
        \centering
        \includegraphics[width=\textwidth]{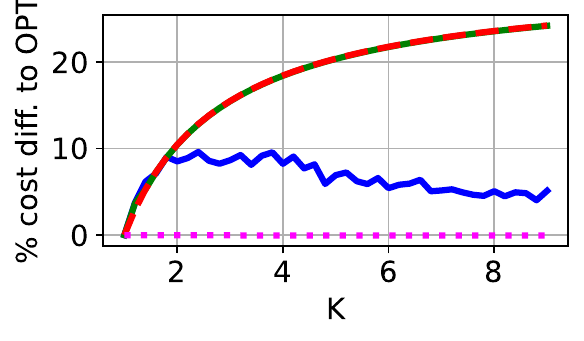}
        %\subcaption{Cost difference to Omniscient for $D=1.67\times L$}\label{fig:gamma_light}
    \end{subfigure}%
    \begin{subfigure}[b]{0.5\columnwidth}
        \centering
        \includegraphics[width=\textwidth]{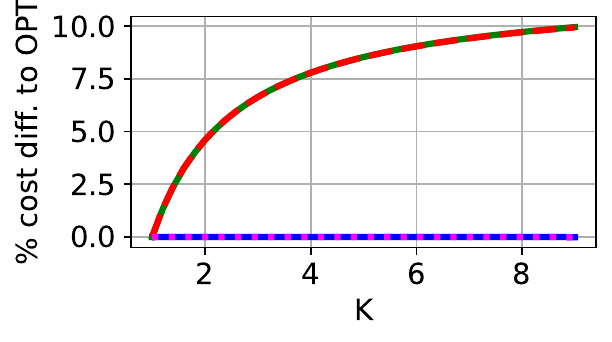}
        %\subcaption{Cost difference to Omniscient for $D=1.1\times L$}\label{fig:gamma_overload}
    \end{subfigure}
    \begin{subfigure}[b]{\columnwidth}
        \centering
        \includegraphics[width=\textwidth]{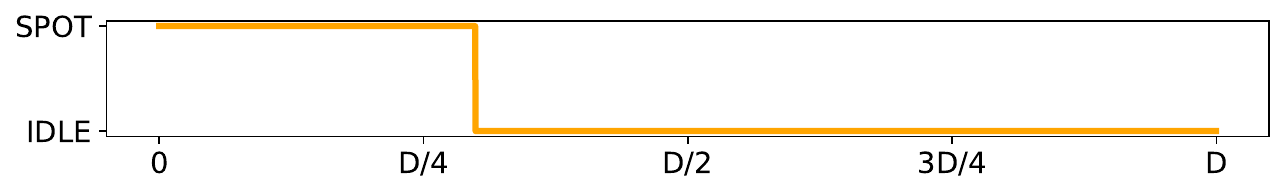}
        %\subcaption{}\label{fig:gamma_overload}
    \end{subfigure}
    \caption{Spotlake Trace Azure D48as-v5-AU-east-1\\First row: \% savings w.r.t on-demand as deadline get stricter.\\ Second row: \% extra cost over \opt{} as $K$ increases. Left plot presents loose deadlines and right plot presents strict deadlines.}
    \label{fig:azure}
\end{figure}

\begin{figure}[t]
\centering
    \begin{subfigure}[b]{\columnwidth}
        \centering
        \includegraphics[width=\textwidth]{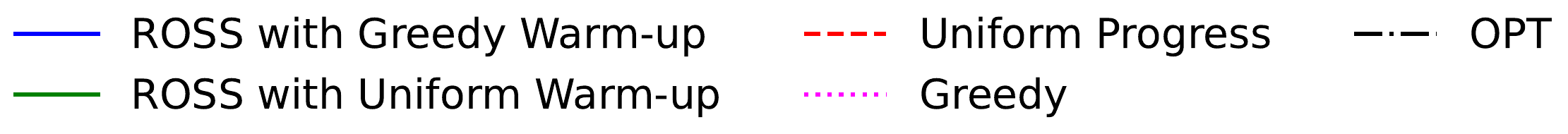}
    \end{subfigure}
    \begin{subfigure}[b]{\columnwidth}
        \centering
        \includegraphics[width=\textwidth]{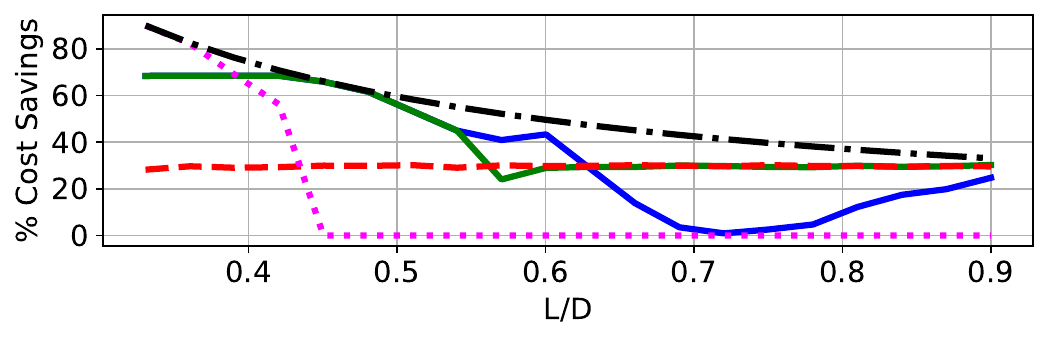}
        %\subcaption{Savings over on-demand}
        \label{fig:exp_light}
    \end{subfigure}
    \vspace{-25pt}
    % \begin{subfigure}[b]{0.5\columnwidth}
    %     \centering
    %     \includegraphics[width=\textwidth]{us-east-1f_v100_1/jf_0.5.pdf}
    %     \subcaption{Cost difference to Omniscient for $D=2\times L$}
    %     \label{fig:exp_overload}
    % \end{subfigure}%
    \\
    \begin{subfigure}[b]{0.5\columnwidth}
        \centering
        \includegraphics[width=\textwidth]{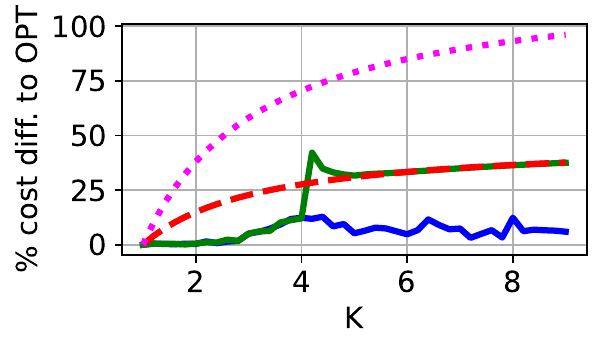}
        %\subcaption{Cost difference to Omniscient for $D=1.67\times L$}\label{fig:gamma_light}
    \end{subfigure}%
    \begin{subfigure}[b]{0.5\columnwidth}
        \centering
        \includegraphics[width=\textwidth]{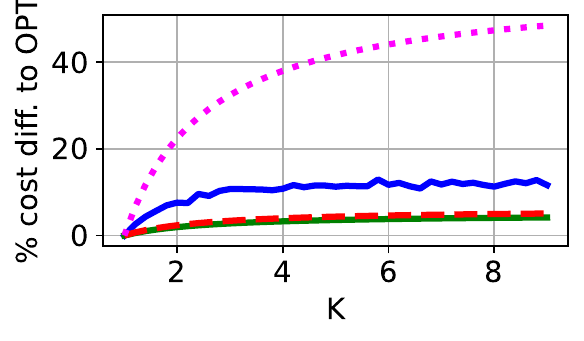}
        %\subcaption{Cost difference to Omniscient for $D=1.1\times L$}\label{fig:gamma_overload}
    \end{subfigure}
    \begin{subfigure}[b]{\columnwidth}
        \centering
        \includegraphics[width=\textwidth]{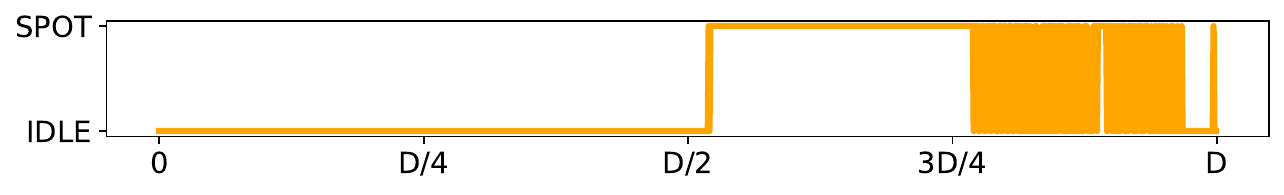}
        %\subcaption{}\label{fig:gamma_overload}
    \end{subfigure}
    \caption{Spotlake Trace AWS-EC2 c3large-us.east-1\\First row: \% savings w.r.t on-demand as deadline get stricter.\\ Second row: \% extra cost over \opt{} as $K$ increases. Left plot presents loose deadlines and right plot presents strict deadlines.}
    \label{fig:aws_lake}
    \vspace{-10pt}
\end{figure}

\begin{figure}[t]
\centering
    \begin{subfigure}[b]{\columnwidth}
        \centering
        \includegraphics[width=\textwidth]{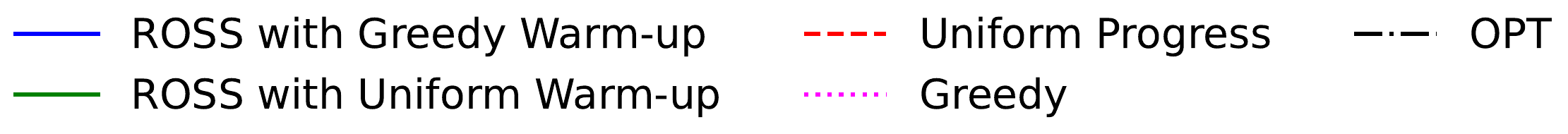}
    \end{subfigure}
    \begin{subfigure}[b]{\columnwidth}
        \centering
        \includegraphics[width=\textwidth]{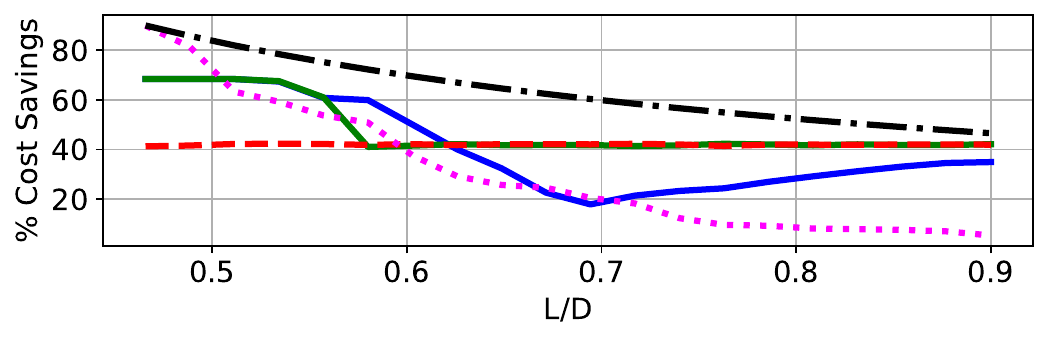}
        %\subcaption{Savings over on-demand}
        \label{fig:exp_light}
    \end{subfigure}%
    \vspace{-15pt}
    \\
    \begin{subfigure}[b]{0.5\columnwidth}
        \centering
        \includegraphics[width=\textwidth]{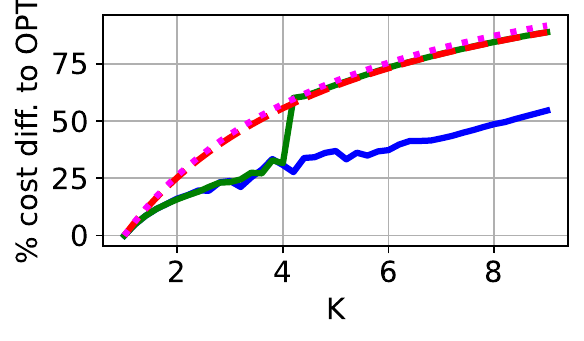}
        %\subcaption{Cost difference to Omniscient for $D = 1.67L$}\label{fig:gamma_light}
    \end{subfigure}%
    \begin{subfigure}[b]{0.5\columnwidth}
        \centering
        \includegraphics[width=\textwidth]{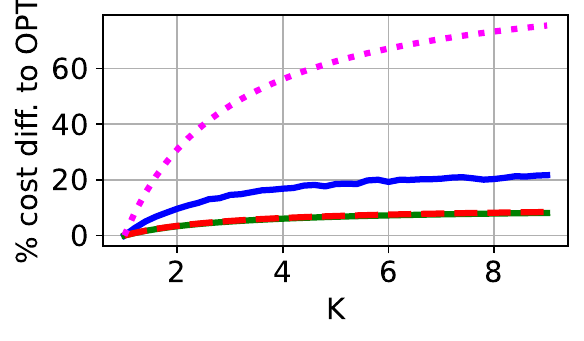}
        %\subcaption{Cost difference to Omniscient for $D = 1.1L$}\label{fig:gamma_overload}
    \end{subfigure}
    \begin{subfigure}[b]{\columnwidth}
        \centering
        \includegraphics[width=\textwidth]{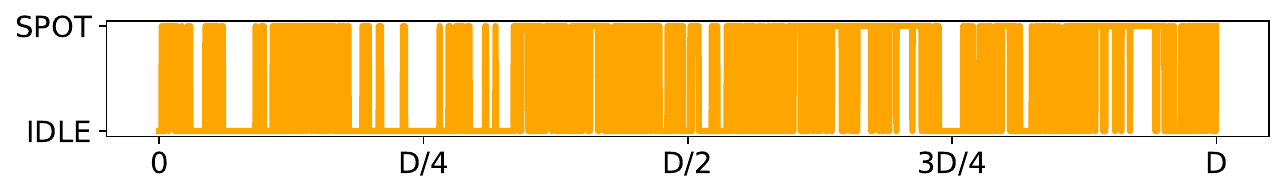}
        %\subcaption{}\label{fig:gamma_overload}
    \end{subfigure}
    \caption{Skypilot Availability Trace us-west-2b-v100-8\\First row: \% savings w.r.t on-demand as deadline get stricter.\\ Second row: \% extra cost over \opt{} as $K$ increases. Left plot presents loose deadlines and right plot presents strict deadlines.}
    \label{fig:aws_sky_avail}
\end{figure}

\begin{figure}[t]
\centering
    \begin{subfigure}[b]{\columnwidth}
        \centering
        \includegraphics[width=\textwidth]{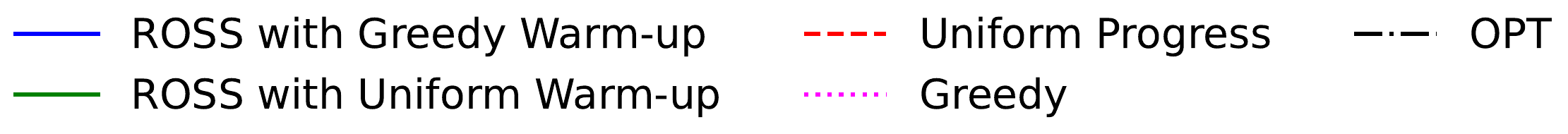}
    \end{subfigure}
    \begin{subfigure}[b]{\columnwidth}
        \centering
        \includegraphics[width=\textwidth]{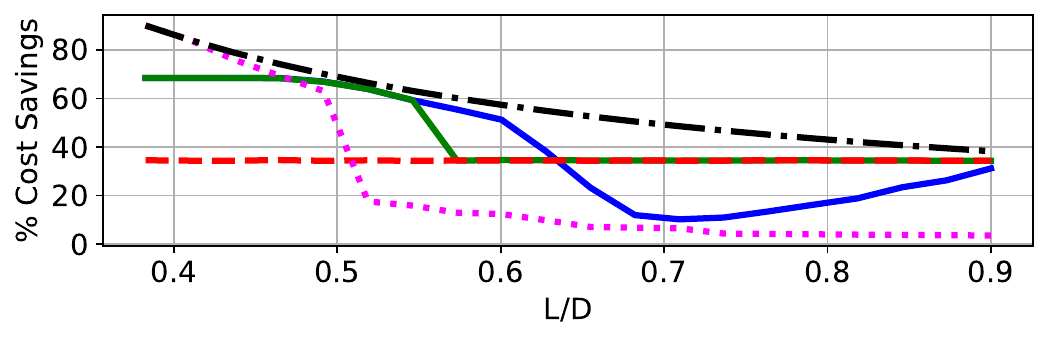}
        %\subcaption{Savings over on-demand}
        \label{fig:exp_light}
    \end{subfigure}
    \vspace{-25pt}
    % \begin{subfigure}[b]{0.5\columnwidth}
    %     \centering
    %     \includegraphics[width=\textwidth]{us-east-1f_v100_1/jf_0.5.pdf}
    %     \subcaption{Cost difference to Omniscient for $D=2\times L$}
    %     \label{fig:exp_overload}
    % \end{subfigure}%
    \\
    \begin{subfigure}[b]{0.5\columnwidth}
        \centering
        \includegraphics[width=\textwidth]{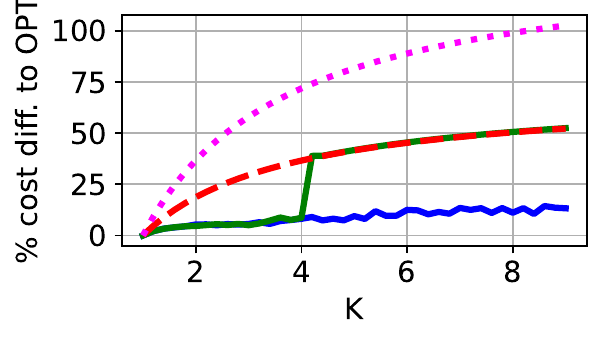}
        %\subcaption{Cost difference to Omniscient for $D=1.67\times L$}\label{fig:gamma_light}
    \end{subfigure}%
    \begin{subfigure}[b]{0.5\columnwidth}
        \centering
        \includegraphics[width=\textwidth]{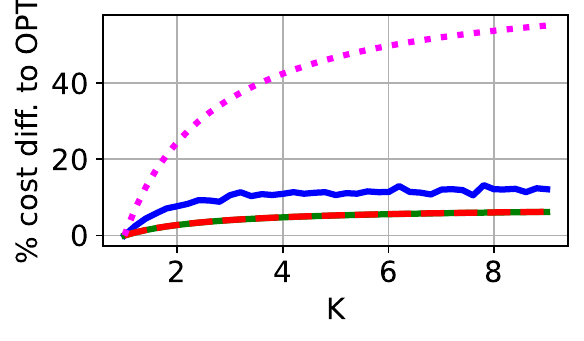}
        %\subcaption{Cost difference to Omniscient for $D=1.1\times L$}\label{fig:gamma_overload}
    \end{subfigure}
    \begin{subfigure}[b]{\columnwidth}
        \centering
        \includegraphics[width=\textwidth]{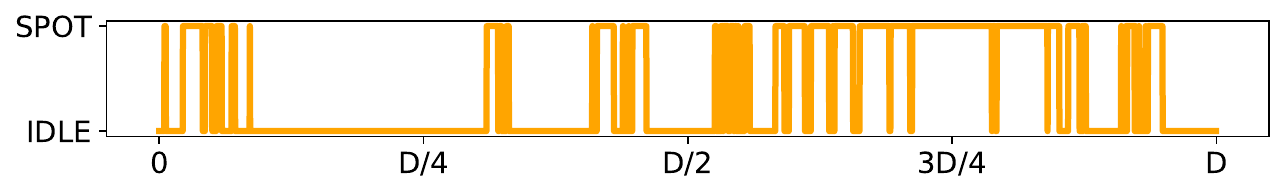}
        %\subcaption{}\label{fig:gamma_overload}
    \end{subfigure}
    \caption{Skypilot Preemption Trace us-east-1f-v100-1\\First row: \% savings w.r.t on-demand as deadline get stricter.\\ Second row: \% extra cost over \opt{} as $K$ increases. Left plot presents loose deadlines and right plot presents strict deadlines.}
    \label{fig:aws_sky_pre}
\end{figure}

Our results reveal that the state-of-the-art Uniform Progress policy, performs poorly under relaxed deadlines—often even worse than the simple greedy approach. In contrast, \ross{} adapts to both loose and tight deadlines by intelligently leveraging available slack and injecting controlled randomization to mitigate worst-case behavior. We now highlight key observations:

\paragraph{\bf Loose Deadlines ($L/D \lessapprox 0.65$)} Uniform Progress performs particularly poorly across all traces in this regime. As shown in Figures~\ref{fig:azure} and~\ref{fig:aws_sky_pre}, it is outperformed by the naive greedy baseline. Figures~\ref{fig:aws_lake} and~\ref{fig:aws_sky_avail} demonstrate that \ross{} consistently outperforms both baselines in the intermediate region ($L/D \in [0.45, 0.65]$). The cost overhead of Uniform Progress with respect to \opt{} increases significantly in favorable conditions, revealing its overly conservative nature. By contrast, the randomization in \ross{} effectively exploits available slack ($D - L$), avoiding the cascade of on-demand costs triggered by greedy scheduling.

\paragraph{\bf Tight Deadlines ($L/D \gtrsim 0.7$)} Greedy scheduling is consistently suboptimal in this regime, as evidenced in Figures~\ref{fig:aws_lake},~\ref{fig:aws_sky_avail}, and~\ref{fig:aws_sky_pre}. Uniform Progress shows more robust performance. However, \ross{}, when initialized with a Uniform Warm-up, closely matches or surpasses Uniform Progress across all traces. Even with a Greedy Warm-up, \ross{} eventually recovers and performs comparably to Uniform Progress for stringent deadlines ($L/D \approx 0.9$), maintaining a gap of less than $10\%$ in both cost savings and overhead to \opt{}.

\paragraph{\bf Impact of Spot Availability Patterns} The four traces represent diverse availability regimes. In high availability scenarios (e.g., Figure~\ref{fig:azure}), Uniform Progress underperforms due to its inherent pessimism. Figure~\ref{fig:aws_sky_pre} shows that under sustained availability, it performs no better than greedy. In contrast, \ross{} effectively adapts—capitalizing on early availability (Fig.~\ref{fig:azure}), abundant spot supply (Fig.~\ref{fig:aws_sky_avail}), and remaining resilient under delayed or sparse spot arrivals (Figs.~\ref{fig:aws_lake},~\ref{fig:aws_sky_pre}).

\medskip
\noindent
In summary, \ross{} consistently balances exploration and exploitation through randomization and adaptive warm-up strategies. When Uniform Progress is overly cautious and greedy is overly aggressive, \ross{} provides a principled and cost-efficient middle ground—achieving the best of both worlds.

\section{Proofs}
\label{sec:proof}
\begin{proof}[\textbf{Proof of Theorem \ref{thm:det_policies}}]
Specifically, the adversary constructs a spot availability pattern in which spot instances appear just after the scheduler commits to renting an on-demand instance, and are preempted immediately whenever the scheduler attempts to switch to a spot instance. Since the algorithm is deterministic, its behavior is fully predictable, allowing the adversary to exploit its decisions. This worst-case construction guarantees that the algorithm incurs high cost, thereby establishing the lower bound.
\end{proof}

\begin{proof}[\textbf{Proof of Theorem 2}]
Consider the following \textbf{class of algorithms} \alg{} that operate in three phases: 
\begin{enumerate}
    \item Phase $1$: For $t\in[0,z)$ where the scheduler randomly chooses an interval of length $\delta$ for on-demand rental while being greedy otherwise and,
    \item Phase $2$: For $t \in [z,T^{NR}_{\alg})$ where \alg{} greedily looks for a spot instance until wait-time slack depletes to zero. 
    \item Phase $3$: For $t\in [T^{NR}_{\alg},D]$ where \alg{} is forced to rent on-demand due to zero wait-time slack and unknown future spot availability,
\end{enumerate}
Any adversary will have the following property:
\begin{enumerate}
    \item Rent \textbf{at most} $\gamma \in [0,1]$ on-demand units,
    \item Supply $\alpha \in [0,z]$ units of spot instance in Phase $1$
    \item Supply $\min\{D-T^{NR}_{\alg},L-\alpha-\gamma\}$ spot instances in Phase 3.
    \item Supply any left-over spot instances in Phase 2, which will amount to
    $$\epsilon = L-(\alpha+\gamma+\min\{D-T^{NR}_{\alg},L-\alpha-\gamma\})$$
\end{enumerate}
Now, \alg{} will pick up all of the $\epsilon$ spot instances in Phase 2. However, the spot instances introduced in Phase 3 can only be caught by the algorithm if
\begin{align}
    L-\alpha-\gamma > D-T^{NR}_{\alg}
\end{align}
because, the \alg{} rents on-demand indefinitely after $T^{NR}_{\alg}$. The above condition coincides with the fact that $\epsilon>0$. This means, if $\epsilon=0$, \alg{} gets NO spot instances in Phase 2 and Phase 3. On the other hand, if $\epsilon>0,$ \alg{} gets ALL spot instances in Phase 2 and 3, while never needing to rent on-demand again. 
% This is because the units of spot instances that will get introduced in Phase 2 and 3, that is, $(L-\alpha-\gamma)$ is larger than the left-over compute for \alg{}, that is $(D-T^{NR}_{\alg}).$

For any random placement of $\delta$ on-demand interval in Phase 1, \alg{} will complete:
\begin{align}
    \delta + \alpha\left(1-\frac{\delta}{z}\right) + \epsilon = \alpha + \delta\left(1-\frac{\alpha}{z}\right) +\epsilon
\end{align}
by the end of Phase 2, which leads to
\begin{align}
    T_{\alg}^{NR} = D-L+\alpha+\delta\left(1-\frac{\alpha}{z}\right)+\epsilon.
\end{align}

% The adversary has no incentive to place spot instances in Phase 2 of \alg{}. The adversary needs at most $1-\gamma$ units of spot instances. Now, Phase 1 and Phase 3 together have a length of
% \begin{align}
%     z+1-\delta
% \end{align}

% The $\gamma$ units of on-demand rental by the adversary can be placed anywhere on the horizon and does make any difference in terms of the competitive ratio.

% % The online algo wants to maximize the usage of spots in Phase 2. This means that the on-demand rental in Phase 1 has to be such that it ca \textbf{afford} to wait till when spots are introduced in Phase 2. 
% Now, the question is whether \alg{} will catch spot instances in Phase 3, based on its $T_{\alg}^{NR}$. The hindsight optimal itself has to finish its own compute load. This means that there exists a point of no-return $T^{NR}_{\adv{}}$ for the \adv{} too, beyond which the \adv{} has to \textit{supply spot instances} to finish its own compute load. $T_{\alg}^{NR}$ and $T_{\adv}^{NR}$ together decide \alg{}'s cost:
Based on outcomes in Phase 2 and 3, the average cost of the algorithm is:
\begin{enumerate}
    \item if $\epsilon>0$, the expected cost of \alg{} is
    \begin{align*}
    &\E[C_{\alg{}}] \\
    &= K\delta + \alpha\left(1-\frac{\delta}{z}\right) + \left( L-\delta-\alpha\left(1-\frac{\delta}{z}\right)\right)\\
    &= L+(K-1)\delta
\end{align*}
\item if $\epsilon=0$, the expected cost of \alg{} is
\begin{align}
    &\E[C_{\alg{}}] \\
    &= K\delta + \alpha\left(1-\frac{\delta}{z}\right) + K\left( L-\delta-\alpha\left(1-\frac{\delta}{z}\right)\right)\\
    &= L+(K-1)\left(L-\alpha\left(1-\frac{\delta}{z}\right) \right)
\end{align}
\end{enumerate}
The cost of the \adv{} is
\begin{align}
    C_{OPT} &= K\gamma + \alpha + (L-\alpha-\gamma)\\
    &= L+(K-1)\gamma
\end{align}
Notice that $\epsilon=0$ only when $D-\alpha-\gamma \leq D-T^{NR}_{\alg}$, which translates to
\begin{align}
    \gamma \geq \delta\left(1-\frac{\alpha}{z}\right) = \gamma^*
\end{align}
Therefore, the cost incurred by \alg{} depends on $\gamma$ as
\begin{align}
    \E[C_{\alg{}}] =\begin{cases}
        L+(K-1)\left(L-\alpha\left(1-\frac{\delta}{z}\right) \right) & \gamma \geq \gamma^*\\
         L+(K-1)\delta & \text{ o.w.}
    \end{cases}
\end{align}
% where
% \begin{align}
%     \gamma^* = \delta\left(1-\frac{\alpha}{z}\right).
% \end{align}
The competitive ratio, therefore, is
\begin{align}
    \frac{L+(K-1)\left(\left(L-\alpha\left(1-\frac{\delta}{z}\right) \right)\mathbbm{1}(\gamma\geq \gamma^*) + \delta\mathbbm{1}(\gamma< \gamma^*) \right) }{L+(K-1)\gamma}
\end{align}
The algorithm \alg{$(z,\delta)$} first decides on a value of $z$ and $\delta$. In response to this, the \adv{$(\alpha,\gamma)$} chooses a value of $\alpha$ and $\gamma.$ This inter-play can be mathematically stated in the following minimax problem:
\begin{align}
    \min_{z,\delta} \max_{\alpha,\gamma} \frac{L+(K-1)\left(\left(L-\alpha\left(1-\frac{\delta}{z}\right) \right)\mathbbm{1}_{\gamma\geq \gamma^*} + \delta\mathbbm{1}_{\gamma< \gamma^*} \right) }{L+(K-1)\gamma}
\end{align}

\textbf{Adversary's Strategies}: \alg{}'s cost has just two values based on \adv{}'s choice of $\gamma$. 
% Hence, \adv{} chooses between:
% \begin{enumerate}
%     \item Strategy $S_1$ (greedy): $\gamma<\gamma^*$, in which case, the competitive ratio is maximized at $\gamma = 0$ or,
%     \item Strategy $S_2$ (judicious): $\gamma \geq\gamma^*$, in which case, competitive ratio is maximized at $\gamma = \gamma^*$
% \end{enumerate}
The adversary will choose one of these two strategies to maximize \alg{}'s competitive ratio.

\noindent \textbf{Strategy $S_1$}: The adversary aims to minimize its cost by only using spot instances for itself. In this case, for any value of $\alpha \in [0,z]$, $T^{NR}_{\alg}>T^{NR}_{\adv} \implies$ \alg{} gets all spot instances in Phase 2. Hence, the cost ratio is $1+(K-1)(\delta/L).$

\noindent\textbf{Strategy $S_2$}: The adversary wants to prevent \alg{} from getting in spot instances in Phase 3, that is after $t=T^{NR}_{\alg}$. This means that the \adv{} must catch up with \alg{} on compute progress, so as to delay spot instances till as late as $T^{NR}_{\alg} \implies \gamma = \gamma^* = \delta\left(1-\frac{\alpha}{z}\right)$.
Now, this leads to the following optimization problem for \adv{} in $\alpha$ and between $S_1,S_2$,
{\scriptsize
\begin{align}
     &\max_{S_1,S_2} \max_{\alpha} \left\{1+(K-1)\frac{\delta}{L}, \frac{L+(K-1)\left(L-\alpha\left(1-\frac{\delta}{z}\right) \right)}{L+(K-1)\delta\left(1-\frac{\alpha}{z}\right)}\right\}\\
     &= \max_{S_1,S_2} \left\{1+(K-1)\frac{\delta}{L}, \max_{\alpha}\frac{L+(K-1)\left(L-\alpha\left(1-\frac{\delta}{z}\right) \right)}{L+(K-1)\delta\left(1-\frac{\alpha}{z}\right)}\right\}
\end{align}
}
% Hence, the scheduler can push the wait for spot instance in Phase 2 till:
% \begin{align}
%     t &= D - \left(1-\delta - \alpha\left(1-\frac{\delta}{z}\right)\right)\\
%     &= y+\delta+\alpha\left(1-\frac{\delta}{z}\right)\\
%     &\geq y+\alpha+\gamma
% \end{align}
% which it wants to be larger than $\alpha+\gamma$ to use entire spot in Phase 2. This gives us
% \begin{align}
%     \delta+\alpha\left(1-\frac{\delta}{z}\right)
%     &\geq \alpha+\gamma\\
%     \delta - \frac{\alpha\delta}{z} \geq \gamma
% \end{align}
% which gives us the condition:
% \begin{align}
%     \delta \geq \frac{\gamma}{1-\frac{\alpha}{z}}
% \end{align}
\textbf{Optimization problem in $\alpha$ for Strategy $S_2$}:
The optimization problem for the adversary is
\begin{align}
    \alpha^* = \arg \max_{0\leq \alpha \leq z} \frac{L+(K-1)\left(L-\alpha\left(1-\frac{\delta}{z}\right) \right)}{L+(K-1)\delta\left(1-\frac{\alpha}{z}\right)}
\end{align}
When solving the optimization problem, what is interesting is that the sign of the derivative of $CR(\alpha)$ w.r.t $\alpha$ is in fact independent of $\alpha$, as shown below. 
% The numerator is the following
% \begin{align}
% \begin{split}
%     -(K-1)&\left(1-\frac{\delta}{z}\right)\left[L+(K-1)\left(\delta-\frac{\alpha\delta}{z}\right)\right]\\
%     &+ \frac{(K-1)\delta}{z}\left[ L+(K-1)\left(L-\alpha\left(1-\frac{\delta}{z}\right)\right) \right]
% \end{split}\\
% \begin{split}
%     = \alpha(K-1&)^2 \left\{\frac{\delta}{z}\left(1-\frac{\delta}{z}\right) - \frac{\delta}{z}\left(1-\frac{\delta}{z}\right) \right\}\\
%     &(K-1)\left\{ KL\frac{\delta}{z} - [L+(K-1)\delta]\left(1-\frac{\delta}{z}\right) \right\}
% \end{split}\\
% = (K-1)&\left\{ KL\frac{\delta}{z} - [L+(K-1)\delta]\left(1-\frac{\delta}{z}\right) \right\}
% \end{align}
% and the denominator is
% \begin{align}
%     \left[L+(K-1)\delta\left(1-\frac{\alpha}{z}\right)\right]^2
% \end{align}
\begin{align}
\frac{d CR(K,\alpha)}{d \alpha} = \frac{(K-1)\left\{ KL\frac{\delta}{z} - [L+(K-1)\delta]\left(1-\frac{\delta}{z}\right) \right\}}{\left[L+(K-1)\delta\left(1-\frac{\alpha}{z}\right)\right]^2}
\end{align}
whose sign is controlled only by the online algorithm $\alg{} = (\delta,z).$ This means that, $CR(\alpha)$ is either always increasing or always decreasing in the range of $\alpha\in [0,z].$ This means that
{\small
\begin{align}
    \max_\alpha &CR(K,\alpha) \\
    &\leq \max\{CR(K,0),CR(K,z)\}\\
    &= \max\left\{ \frac{K}{1+(K-1)\frac{\delta}{L}}, 1+(K-1)\left(1+\frac{\delta}{L}-\frac{z}{L}\right) \right\}
\end{align}
}
% To start with, we can set $z=1$ and $\delta = \frac{\sqrt{K}-1}{K-1} \in (0,1)$. This leads to
% \begin{align}
%     \frac{K}{1+(K-1)\delta} = \frac{K}{\sqrt{K}} &= \sqrt{K}\\
%     1+(K-1)(1+\delta-y-z) = 1+\sqrt{K}-1 &= \sqrt{K}
% \end{align}
% which gives us an upper bound on the worst-case performance,
% \begin{align}
%     CR_{\alg}\leq \sqrt{K}
% \end{align}
% with $\delta = \frac{\sqrt{K}-1}{K-1}$ and $z = 1-y.$

\noindent\textbf{Algorithm's Strategies}:\\
\noindent \textit{Against Strategy $S_1$}: Against a greedy adversary, the algorithm can simply make $\delta \to 0$ to obtain a cost ratio of $1$. However, this is sub-optimal as we will see below in Strategy $S_2.$

\noindent \textit{Against Strategy $S_2$}: 
Recall that $\frac{d CR(K,\alpha)}{d\alpha}$ is either positive or negative. Based on this, adversary can supply $\alpha =0$ or $\alpha = z$, leading to a bad competitive ratio for naive choices of $\delta$. 
% Now, the question is, what the algorithm can do against Strategy $S_2$. To see this, we need to understand how the adversary controls $CR(\alpha)$. The following table shows naive values of $\delta$, in response to which \adv{} manages to force a poor competitive ratio. 
\begin{table}[ht]
\centering
\setlength{\extrarowheight}{2pt}
\begin{tabular}{cc|c|c|}
  & \multicolumn{1}{c}{} & \multicolumn{2}{c}{$\substack{\text{Adversary's decision}\\\text{spot suppply }(\alpha)}$}\\
  & \multicolumn{1}{c}{} & \multicolumn{1}{c}{$0$}  & \multicolumn{1}{c}{$L$} \\\cline{3-4}
  \multirow{2}*{$\substack{\text{Scheduler's decision}\\\text{On-demand length }(\delta)}$}  & $0$ & \red{$K$} & $1$ \\\cline{3-4}
  & $L$ & $1$ & \red{$K$} \\\cline{3-4}
\end{tabular}
\end{table}

\noindent However, if we make 
\begin{align}
    \frac{d CR(K,\alpha)}{d\alpha} = 0
\end{align}
it will impair the adversary from doing any policy-specific damage. From the above calculations, it can be verified this condition is same as
\begin{align}
    CR(K,0) = CR(K,z).
\end{align}
% For, $z = 1-y$, $\delta = \frac{\sqrt{K}-1}{K-1}$ leads to the above adversarial-optimal condition (w.r.t $\delta$), providing improved performance.

% To extract the last bit of performance, we need to perform the minimization of CR w.r.t $z \in [-y,1]$. 
Putting $\delta = \rho z$ where $\rho \in [0,1]$ and solving for $CR(K,0) - CR(K,z)=0$ or $\frac{d CR(K,\alpha)}{d\alpha}=0$, we get
\begin{align}
    \rho^2 + \rho\left(\frac{K+1}{(K-1)(z/L)}-1\right)-\frac{1}{(K-1)(z/L)} = 0.
\end{align}
The positive root for this quadratic is
\begin{align}
    \rho^* =& -\frac{1}{2}\left(\frac{K+1}{(K-1)(z/L)}-1\right) \\
    &+ \frac{1}{2}\sqrt{\left(\frac{K+1}{(K-1)(z/L)}-1\right)^2 + \frac{4}{(K-1)(z/L)}}.
\end{align}
which gives the optimal $\delta^*$ for Strategy $S_2$ as:
\begin{align}
    \delta^*(z) =& -\frac{1}{2}\left(L\frac{K+1}{K-1}-z\right) \\
    &+ \frac{1}{2}\sqrt{\left(L\frac{K+1}{K-1}-z\right)^2 + \frac{4zL}{K-1}}
\end{align}
Since  this value of $\delta^*(z)$ leads to $CR(K,0) = CR(K,z)$, we will use the former expression, that is $\frac{K}{1+(K-1)\frac{\delta^*(z)}{L}}$.\\

\noindent\textbf{Handing $S_1$ and $S_2$ together}:
Finally, the \ross{} has to decide on the optimal value of $z$ against strategy $S_2$. Notice that in the range of $z \in [0,D]$, $\delta^*(z)$ also covers the optimal $\delta$ against Strategy $1$, that is, zero with $z = 0.$ This means that the optimal choice of $z$ can be made to encompass both strategies, $S_1, S_2$ in the following manner:
\begin{align}
    \min_{z\in[0,D]} \max \left\{1+(K-1)\frac{\delta^*(z)}{L},\frac{K}{1+(K-1)\frac{\delta^*(z)}{L}} \right\}
\end{align}
Now, $\delta^*(z)$ is a strictly increasing function in $z$. This means $\max \left\{1+(K-1)\frac{\delta^*(z)}{L},\frac{K}{1+(K-1)\frac{\delta^*(z)}{L}} \right\}$ is minimized when the terms inside are equal:
\begin{align}
    1+(K-1)\frac{\delta^*(z)}{L} = \sqrt{K}
    \implies \delta^*(z^*) = \frac{L}{\sqrt{K}+1}.
\end{align}
For this value of $\delta^*(z^*)$, solving 
\begin{align}
\begin{split}
    \frac{L}{\sqrt{K}+1} =& -\frac{1}{2}\left(L\frac{K+1}{K-1}-z\right) \\
    &+ \frac{1}{2}\sqrt{\left(L\frac{K+1}{K-1}-z\right)^2 + \frac{4zL}{K-1}}
\end{split}
\end{align}
gives $$z^* = L.$$

Under Strategy $S_1$ $(\gamma=0)$, \ross{} is always ahead of \adv{} and consequently rents on-demand for at most $\delta$ units. Therefore, the cost ratio is always $1+(K-1)\delta/L.$

Now, for Strategy $S_2$, there is a phase shift in how \ross{} behaves for small $D$. For the three phases in \alg{} to exist,
\begin{align*}
    z\leq T^{NR}_{\alg} = D-L+\delta+\alpha\left(1-\frac{\delta}{z}\right),
\end{align*}
where $T^{NR}_{\alg}$ increases with $\alpha.$ Now, previously, we had set $z^*=L$ and $\delta^* = \frac{L}{\sqrt{K}+1}.$ \alg{} is valid for any unknown $\alpha\geq 0$ when $z^*\leq D-L+\delta + 0$, that is
\begin{align}
    D\geq 2L-\delta = \frac{1+2\sqrt{K}}{1+\sqrt{K}}L.
\end{align}
For, $D\leq\frac{1+2\sqrt{K}}{1+\sqrt{K}}L,$ Phase 1 cannot be of length $L$. Therefore, the length of Phase $1$ can be at most:
\begin{align}
    z = D-L+\delta
\end{align}
where $\delta$ is yet to be decided. Recall that the competitive ratio is for Phase $1$ of general length $z\geq 0$,
\begin{align}
\begin{split}
    CR = \max\bigg\{1+(K&-1)\frac{\delta}{L},\frac{K}{1+(K-1)\frac{\delta}{L}},\\
    &1+(K-1)\left(1+\frac{\delta}{L}-\frac{z}{L}\right) \bigg\}
\end{split}
\end{align}
which is minimized at the maximum value of $z$, meaning
\begin{align}
\begin{split}
    CR = \max\bigg\{1+&(K-1)\frac{\delta}{L},\frac{K}{1+(K-1)\frac{\delta}{L}},\\
    &1+(K-1)\left(2-\frac{D}{L}\right) \bigg\}
\end{split}
\end{align}
Keeping $\delta = \frac{L}{1+\sqrt{K}}$ and $z = \min\{D-L+\delta,L\}$, we have 
\begin{align}
    CR_{\ross{}} =\begin{cases}
        \sqrt{K} & D\geq \frac{1+2\sqrt{K}}{1+\sqrt{K}}L\\
        1+(K-1)\left(2-\frac{D}{L}\right) & D\leq \frac{1+2\sqrt{K}}{1+\sqrt{K}}L
    \end{cases}
\end{align}
\end{proof}

\begin{proof}[\textbf{Proof of Theorem 3}]
The proof of this result follows a fluid model. Consider the horizon to be from the origin to the deadline, with any timestamp on it denoted as $t \in [0,D].$ We can model any online algorithm \alg{} using two parameters:
\begin{enumerate}
    \item $p(t)$: the probability of \alg{} choosing an on-demand instance at time $t$
    \item $T^{NR}_{\alg}$: the point of no-return, starting which $p(t) = 1$ $\forall$ $t\geq T^{NR}_{\alg}$
\end{enumerate}
Any online algorithm can have its choice of $p:[0,D]\to [0,1]$. But in addition it can also control $T^{NR}_{\alg} \in [0,D]$ under any spot-arrival pattern by pushing it as far as it wants through additional on-demand rental (and perhaps suffering from higher cost).

Any adversarial instance, on the other hand, can be defined by the following two quantities:
\begin{enumerate}
    \item $\alpha(t)$: Probability of supplying a spot instance at time $t$
    \item $\gamma^*$: Total on-demand work it does in $[0,D].$
\end{enumerate}
As discussed in the previous section, the adversary has an incentive to rent on-demand $(\gamma^*>0)$ to catch up to $\alg{}$ and prevent it from getting any spot instances at $t=T^{NR}_{\alg}$, post which \adv{} can supply spot instances. So, either $\gamma^* = 0$ (greedy strategy $S_1$) or \adv{} can match \alg{} (Strategy $S_2$). 
%\red{(Refer to Section II(A): Adversary's strategies)}

Now, progress made by \alg{} till the point of no-return is
\begin{align}
    &\int_0^{T^{NR}_{\alg}} p(t)dt + \int_{0}^{T^{NR}_{\alg}} \alpha(t)(1-p(t))dt\\
    &= \int_0^{T^{NR}_{\alg}} \alpha(t)dt + \int_{0}^{T^{NR}_{\alg}} p(t)(1-\alpha(t))dt\\
    &\geq \int_0^{T^{NR}_{\alg}} \alpha(t)dt \label{eqn1}
\end{align}
Under Strategy $S_1$, \adv{} has a total cost of $L$. Now, \adv{} itself has to complete the compute by the deadline, implying $\int_0^{D} \alpha(t) = 1$. From this and \eqref{eqn1}, it is clear that \alg{} will always be ahead of \adv{} in terms of running progress, meaning spots instances supplied by \adv{} will always be caught by \alg{}. Consequently, \alg{} will end up doing an on-demand work of just $\int_0^{D}p(t)dt$. Hence, cost ratio under Strategy $S_1$ is
\begin{align}
    CR^{(1)}_{\alg} = 1+(K-1)\frac{1}{L}\int_0^{D}p(t)dt
\end{align}
On the other hand, to perform Strategy $S_2$, \adv{} has to perform
\begin{align}
    \gamma^* = \int_{0}^{T^{NR}_{\alg}} p(t)(1-\alpha(t))dt
\end{align}
amount of on-demand work to qualify for supplying spot instances just after $T^{NR}_{\alg}$. In this case, cost of any \alg{} is
\begin{align}
\begin{split}
    &\E[C_{\alg}]\\
    &= K\int_0^{T^{NR}_{\alg}} p(t)dt + \int_{0}^{T^{NR}_{\alg}} \alpha(t)(1-p(t))dt \\
    &+ K\left(L-\int_0^{T^{NR}_{\alg}} p(t)dt - \int_{0}^{T^{NR}_{\alg}} \alpha(t)(1-p(t))dt \right)
\end{split}\\
\begin{split}
    =& L+ (K-1)\left(L - \int_{0}^{T^{NR}_{\alg}} \alpha(t)(1-p(t))dt \right)
\end{split}
\end{align}
This will lead to a cost ratio
\begin{align}
    CR^{(2)}_{\alg} = \frac{L+ (K-1)\left(L - \int_{0}^{T^{NR}_{\alg}} \alpha(t)(1-p(t))dt \right)}{L+(K-1)\int_{0}^{T^{NR}_{\alg}} p(t)(1-\alpha(t))dt}
\end{align}
The competitive ratio is the maximum between the cost ratios for Strategies $S_1$ and $S_2$. Hence,
\begin{align}
    CR_{\alg} &= \max_{\alpha(t):[0,D]\to[0,1]} \max\left\{CR^{(1)}_{\alg},CR^{(2)}_{\alg}\right\}\\
    &= \max\left\{CR^{(1)}_{\alg}, \max_{\alpha(t):[0,D]\to[0,1]}CR^{(2)}_{\alg}\right\}
\end{align}
Now, we can lower bound the second term by using the specific $\alpha(t) = 0$, that is
\begin{align}
    \max_{\alpha(t):[0,D]\to[0,1]}CR^{(2)}_{\alg} &\geq \frac{KL}{L+(K-1)\int_{0}^{T^{NR}_{\alg}} p(t) dt}\\
    &\geq \frac{K}{1+(K-1)\frac{1}{L}\int_{0}^{D} p(t)dt}.
\end{align}
and further lower bounding it using $T^{NR}_{\alg} \leq D$. Putting everything together,
\begin{align}
    CR_{\alg} &\geq \max\left\{ CR^{(1)}_{\alg}, \frac{K}{CR^{(1)}_{\alg}}\right\}\\
    &\geq \sqrt{K}
\end{align}
We now show that the competitive ratio for $D\leq \frac{1+2\sqrt{K}}{1+\sqrt{K}}L$ is in fact tight. 

The class of algorithms \alg{} decide on-demand commitment for the horizon $p(t):[0,D]\to[0,1]$ without knowledge of future spot-availability $\alpha(t):[0,D]\to [0,1]$. Any such algorithm has to make a decision on a timestamp $T$, after which it has to stop being random, that is $$p(t) = 1 \text{ } \forall \text{ } t\geq T.$$ This is because \alg{} has to choose on-demand with probability $1$ starting the point of no-return $T^{NR}_{\alg}$,
\begin{align}
    p(t) = 1 \text{ } \forall \text{ } t\geq T^{NR}_{\alg},
\end{align}
and consequently adhere to,
\begin{align}
    T\leq T^{NR}_{\alg}.
\end{align}
Now, the point of no-return is the timestamp where left-over progress equals left-over time to deadline,
\begin{align}
\begin{split}
    &D-T^{NR}_{\alg}(\alpha) \\
    &= L-\int_0^{T^{NR}_{\alg}(\alpha)} p(t)dt - \int_0^{T^{NR}_{\alg}(\alpha)}\alpha(t)(1-p(t))dt
\end{split}
\end{align}
meaning
\begin{align}
\begin{split}
    &T^{NR}_{\alg}(\alpha)\\
    &= D-L+\int_0^{T^{NR}_{\alg}(\alpha)} p(t)dt + \int_0^{T^{NR}_{\alg}(\alpha)}\alpha(t)(1-p(t))dt
\end{split}
\end{align}
Notice that $T^{NR}_{\alg}(\alpha)$ is a function of $\alpha(\cdot)$ which \alg{} does not know when deciding $T.$ \alg{}, hence, has to ensure 
\begin{align}
    T\leq T^{NR}_{\alg}(\alpha) \text{ } \forall \text{ } \alpha(\cdot):[0,D]\to[0,1].
\end{align}
Since, $T^{NR}_{\alg}(\alpha)$ increases with $\alpha(\cdot)$, \alg{} has to decide on a $T$ that satisfies:
\begin{align}
    T \leq T^{NR}_{\alg}(\mathbf{0}) = D-L+\int_0^{T^{NR}_{\alg}(\alpha)} p(t)dt
\end{align}
which translates to
\begin{align}\label{eqn2}
    T\leq D-L+\int_0^{T} p(t)dt
\end{align}
as $p(t) = 0$ $\forall$ $t\in [T,T^{NR}_{\alg}(\mathbf{0})].$ Now, the strategy of the \adv{} involves matching \alg{} up to $t=T$ entirely using spot instances, that is $$\alpha(t) = 1 \text{ } \forall \text{ } t \in [0,T]$$
then pulling the plug on spot instances immediately. 

\alg{} can only be deterministic in $[T,D]$. And whatever \alg{} does, \adv{} matches using spot instances, as it knows to place them exactly when \alg{} is choosing on-demand (not being greedy/waiting for spot). Therefore, the only spot instances \alg{} receives in during its random phase, $[0,T]$ and its cost is
\begin{align}
    \E[C_{\alg}] &= L+(K-1)\left(L-\int_0^T \alpha(t)(1-p(t))dt \right)\\
    &= L+(K-1)\left(L-\int_0^T (1-p(t))dt \right)\\
    &\geq 1+(K-1)(2L-D)
\end{align}
where the last inequality is from \eqref{eqn2}. The cost of \adv{} is always $1$ as it runs only on spot instances, before and after $t=T.$ Consequently, competitive ratio is lower bounded as
\begin{align}
    CR_{\alg} \geq 1+(K-1)\left(2-\frac{D}{L}\right)
\end{align}

\end{proof}

\section{Final Thoughts}
In this work, we have made several key advances in the area of online scheduling of delay-sensitive cloud jobs. First, we establish that existing deterministic policies are inherently limited with a worst-case competitive ratio of $\Omega(K)$. Second, our proposed \ross{} algorithm achieves a competitive ratio that is substantially better $(\sqrt{K})$, under moderate deadline slack, than existing approaches, and this performance is proven to be optimal via a matching lower bound. Lastly, our empirical study using public spot trace datasets validates the theoretical benefits of \ross{}, demonstrating its cost-effectiveness and robust performance under both loose and tight deadline scenarios. 
% \newpage

% \section{Diff wrt $\alpha$}
% We consider the expression
% \[
% f(\alpha) = \frac{1 + (K-1)\left(1 - \alpha \left(1 - \frac{\delta}{y + z}\right)\right)}{1 + (K-1)\delta \left(1 - \frac{\alpha}{y + z}\right)}.
% \]
% We rewrite it in a compact form as
% \[
% f(\alpha) = \frac{A - B\alpha}{C - D\alpha},
% \]
% where
% \[
% A = 1 + (K-1) = K, \quad B = (K-1)\left(1 - \frac{\delta}{y + z}\right),
% \]
% \[
% C = 1 + (K-1)\delta, \quad D = (K-1)\delta \cdot \frac{1}{y + z}.
% \]
% Taking the derivative of \( f(\alpha) \) with respect to \( \alpha \) and applying the quotient rule, we have
% \[
% f'(\alpha) = \frac{-B(C - D\alpha) - (A - B\alpha)(-D)}{(C - D\alpha)^2}.
% \]

% \[
% \boxed{ f'(\alpha) = \frac{ A D - B C }{ \left( C - D \alpha \right)^2 } }
% \]
% This implies that the sign of the derivative with respect to $\alpha$ depends on sign of  \( AD - BC \) can be written in quadratic form with respect to \( \delta \) as
% \[
% AD - BC = \frac{K - 1}{y + z} \left[ (K - 1)\delta^2 + \left( (K + 1) - (K - 1)(y + z) \right)\delta - (y + z) \right].
% \]
% Explicitly, we have
% \[
% AD - BC = \frac{K - 1}{y + z} \left[ a \delta^2 + b \delta + c \right],
% \]
% where
% \[
% a = K - 1, \quad b = (K + 1) - (K - 1)(y + z), \quad c = - (y + z).\]
%\clearpage
\bibliographystyle{ieeetr}
\bibliography{refs}
\end{document}